# A Physical Perspective to Human Migration Phenomenon

Mina Ghanbari[1], Ghader Rezazadeh[*2,3]

[1]Mechanical Engineering Department, Urmia University of Technology, Urmia, Iran
[2]Mechanical Engineering Department, Urmia University, Urmia, Iran
[3]South Ural State University, Chelyabinsk, Russian Federation.

**Abstract**
This paper presents a physical perspective on the human migration phenomenon claiming that human social behaviors are somehow but not completely inspired by nature. Human displacement or migration in the world is highly affected by equivalent weighted wealth and multidimensional poverty distribution across countries. In this work, it has been shown that the equivalent weighted wealth as an important characteristic of a society can be introduced in a mathematical language in terms of a country gross productivity factor, safety and social factors, environmental and economic factors, political and healthy factors, educational and cultural factors and so on using some weights, where these weights can be determined using regression or machine learning techniques. By inspiring from the natural phenomena such as heat conduction models; heat conduction and heat convection models, a constitutive model for migration flow rate in terms of the equivalent weighted wealth gradient descent has been given where accompanying with other balance laws it can make possible to determine wealth distribution and migration flow rate across the world. Some critical conditions such as earthquakes, war, famine, insecurity, and more prevent the natural transmission of wealth and lead to the intensification of non-uniformity in wealth distribution. The increase in the equivalent weighted wealth difference between countries leads to an enhanced legal migration rate as well as a boosted illegal migration flow rate where it exceeds *a threshold value.* The introduced nature-inspired model could significantly predict wealth distribution as well as migration rate, which will have helped those who try to have a world with a logical almost uniform equivalent weighted wealth distribution and will have made the world a better place for all of the people.

1. **Introduction**
**1.1 General definition of the immigration phenomenon**
The immigration phenomenon is defined as the international movement of people into a foreign country to dwell there permanently or temporarily. Immigration can be voluntary as people who forsake their homes in search of high-wage jobs as well as regions with better environmental conditions, and not voluntarily as refugees that are forced to escape their origin country due to fear of being arrested and persecuted on their religion, political beliefs or membership in a particular social group works. The immigration phenomenon can be legal or illegal. In legitimate immigration, the immigrants are accepted for one of the reasons needed to enter as a family, employment, humanitarian, or visa lottery. In illegal immigration, people enter the country via the borders in a variety of ways without the government allowance.

In general, the basic or essential factors of immigration, which encourage people to vacate their homes and select another destination for settling and working there, are called push and pull factors. Push factors represent the factors existing in the immigrants' origin country and seem to be the reasons which persuade and sometimes oblige people to abandon their country. The factors existing in the modern country, which encourage the people to migrate to that country, are considered the pull factors. Safety, environmental, economic, social, and political factors are examples of push and pull factors. Among the safety factors, the quarrel comprises the most significant factor causing forced immigration around the world. Persecution and harassment based on political beliefs and religion encourage people to migrate to a country where they can have freedom and safety in their lives. In 2016, the recording of thousands of crimes in Guatemala, El Salvador, and Honduras caused these countries to remain the most severe and vigorous regions in the world. These conditions lead to the migration of people to secure regions. Recently, the Rohingya refugee crisis in Myanmar with mass illegal migration of their Muslim population, to southern Bangladesh has captured the attention of the world.

Environmental factors are the other substantial factors that have caused immigration in recent years, significantly. Water shortage, which leads to crop failure, food leakage, and reduction of agricultural jobs, abet people to relocate to other places with better environmental conditions where they can find more job positions. Water, air, and soil pollutions due to creating a considerable risk of health can also cause people to escape to locations with better air quality. Natural disasters such as earthquakes and tsunamis are the other environmental factors that may lead to the migration of people. The main reasons for the human movement in economic migration are mostly lack of job positions and inadequate wages. The mentioned reasons cause people to migrate to areas where they can find jobs with higher wages. Social factors that encourage migration originate from a human need to ensure better opportunities for themselves and their families.

## 1.2. Theories and approaches to modeling migration dynamics

Models of human migration dynamics afford tenacious tools to estimate the flow of migrants, determine the effect of a policy, measure the cost of political and physical resistances, and more. In conventional theories on migration, the migration flows are described because of wage differentials in joblessness levels. These early migration theories cannot explain the migration flows in small size and the presence of vast return-migrations. Most of the migration theories assume that migrations are permanent though it is known that many migrations are temporary or repetitive. However, in theories that by modeling timing departure, the temporal nature of migration has been taken into account, assume that the migrants are not permanent and they leave the host country at the end (Goldstein 1964; Massey 1988; Duleep 1994; Dustmann 1995; Dustmann 2000; Constant and Massey 2003). In one of the studies on return migration, the optimal migration duration was analyzed (Dustmann 2002). In the proposed life-cycle theory, the decision of the migrants to return to their home country was intellectualized despite having continuously higher stipends in the host country. In other work, the temporary forms of migration were discussed in detail by Dustmann, and Weiss (2007). In that work, three motives for a temporary migration were considered. The complementarities between expenditure and the location where expenditure takes place, differences in relative prices in the host and home country, and the possibility of gathering human capital abroad increase the immigrant's revenue potential back home.

Duration models or event history models have also been used for modeling migration dynamics.

Detang-Dessendre and Malho (2007) presented a model estimated by the distance of the move. In that model, the effects of different employment-status transitions on migration choices were considered from a theoretic perspective. The results showed that the long-distance migration hazard related to labor market variables is considerably higher among job-gainers compared to the other transition groups.

In conventional duration models of migration, it is assumed that all migrants are movers. This means that all immigrants leave the host country eventually. However, some of the immigrants may never leave. To take into account the possibility that some of the migrants are temporary and some are permanent, the approach of mover-stayer was developed by Boag (1949) and was applied by Dunsmuir et al. (1989) for the labor market transitions. In the dynamic model presented by Nikonov and Tarasyev, the dynamics of labor force movement between Russian regions in continuous time was described [12]. The modified presented model could perfectly demonstrate the dynamics of migration processes and their effects on regional economics. They indicated the results of migration modeling in conditions of a closed system on a graph; therefore, the dependency between total labor force, salaries levels, and the number of vacancies and unemployed people in several regions could be traced. Molho (1984) constructed a dynamic model of migration in Great Britain based on time-series data. By applying a two-stream gravity model, regional push and pull factors were derived based on the revealed preferences of migrants. The obtained results suggested many interesting dynamic responses, implying lengthy and important lags in the estimated relationships. Mason (2017) presented a dynamic model in which citizens of a country vulnerable to damages from the change of climate may move to a second country, from which a steady stream of greenhouse gases occurs. It was shown that in the presence of migration the long-run carbon stock and time path of production is smaller. As the average wage significantly affects the migration processes between the origin country and destination, Tarasyev et al. (2019) presented a model in which the feedback of the average wage determined by the supply of labor resources was taken into account. Further, in this model the influence of the age distribution of migration flows on the labor market in the host country was considered. Hence, the potential of migration flows and their impact on the socio-economic system could be evaluated.

In modern migration theories, the prevailing theory in illustrating the cause of migration is the neoclassical theory. It underlies assumptions that migration is motivated primarily by logical economic consideration of relative financial benefits and costs (Todaro 2006). This theory claims that migration occurs due to differentials of wage and income and the probability of employment. As it predicts a linear relationship between the wage differentials and migration flow, it is unable to explain the migration ceases before equality of wage differentials (Krieger and Bertrand 2006).

The new economics theory of migration is another theory, which claims that the decision of migrants is affected by a general set of factors shaped in the home country. It is subjected to a critique that has limited applicability. It means that it is difficult to isolate the effect of market imperfections in migration decisions from other variables as wage and employment.

Sirojudin (2009) reviewed the economic theories of migration. Economy theories of rational choice, supply and demand, world system, and network are features of economic theories. In rational choice theory for emigration, the costs and the benefits are weighted therefore can help recognize the personal and household considerations for emigration. The process of rational choice consists of the scrutiny of the factors as the differences of the living cost between the home and host country, economic and political changes, differences in wages of the laborers, etc (Sirojudin 2009). There is a challenge to the interpretive power of the rational choice theory and that is the role of the family. Messy et al. (1993) showed that the push factors for emigration are in addition to increasing the income-earning it is also for managing risk in dealing with the potential of market failure. In the rational choice theory, enough attention has not been paid to the structural factors that can affect family and individual behaviors. In supply-and-demand theory, the demand for labor in developed industrial countries is considered as a strong pull factor and the oversupply of labor as a strong push factor. Vujicic et al. (2004) showed that migration of laborers happens in the case of the migration of health care specialists from developing countries to Canada and the United States and can be an example of supply and demand dynamics.

In the world system theory, the determinant of migration and the structural changes induced by the flow of capital are linked. In theory, migration is a function of globalization, so it can be applicable at the global level. In other words, in world-system theory, the complex political and economic factors are emphasized that create a strong push to emigrate by exploiting third-world countries. Skeldon (2002) introduced international migration as a function of the economic differences between countries. He claims that the actual economic disparities as well as the relative poverty, rather than absolute poverty lead to international migration. Synthesizing the evidence on the risk of infection and transmission among migrants, refugees, asylum seekers, and internally displaced populations by Hintermeier et al. (2021). The dynamics of immigrant communities in both overseas countries and countries of origin can be explained by network theory. As networks create the social structures required to maintain the migration process, it makes the process self-sustaining (Constant and Massey 2002). Taylor (1999) recognized the growth potential of remittances from a NELM perspective and referred to the empirical evidence that remittances may be a positive factor in the economic growth of a country.

Modeling this migration phenomenon mathematically based on the Earth's gravity force is another approach in migration dynamics. Various models have been proposed for human migration utilizing gravity force, which is considered as one of the commonly-used models for predicting gross migration flows. It considers the movement of people like the interaction between two earthy units as countries and is derived from Newton's law of gravity. Its advantages including the goodness of fit in most applications, sensorial consistency with other migration theories, and ease of determination have made it a beneficial theory compared to other migration theories. Many types of research have been conducted on developing gravity modeling of migration and overcoming its shortcomings. Beine et al. (2016) introduced the theoretical foundations of the gravity model for estimating international migration. Further, the basic difficulties that have to be acquired in the econometric analysis were discussed and reviewed in this work. The development of the gravity immigration model was proposed by Lewer and Berg (2008). Testing the model showed its usefulness with high illustrative power. The influence of demographic, economic, and political factors on the composition and size of migration flows to North America was investigated by Karemera et al. (2000). They adjusted a gravity model in which immigration characteristics and regulations specific to the destination and origin countries were taken into account. Malaj and Rubertis (2017) studied the determinants of emigration from the Western Balkans utilizing the gravity model. In that work, they concluded that in the gravity model, unemployment, the standard of living, and corruption are three significant parameters that are related to the concerns for the Western Balkans. Kim and Cohen (2010) determined international migration flows to industrial countries by proposing a new approach beyond gravity. Curiel et al. (2018) found the city-to-city migration follow scaling laws were founded to analyze the individuals' migration from and to cities in the US. It was observed that the size of both home and host cities plays a significant role in deciding the individuals to select their destination. They showed that the migration of individuals from similar-sized cities is occurring more frequently. However, individuals from large cities do not tend to migrate but if they do, they tend to move to other large cities. Constructing a gravity-scaling-based model considering the impact of distance showed that the scaling laws are an important feature of human migration that can dominate the limits of the radiation and the gravity models of human migration. Bunea (2012) using county data for the period 2004-2008 and applying the gravity model explored the potential determinants of internal migration in Romania. The results indicated that from a static point of view real gross product per capita, population size, road density, amenity index, and crime rate are key factors of migration. From a dynamic point of view, impacts of population size and amenity index, and previous migration ratio are very significant in determining the migration flow.

**1.3 New perspective on the migration phenomenon**
Beyond the theories and approaches mentioned in the literature review, this paper presents a novel physical perspective on the migration phenomenon where the social behavior of humans as their displacement between countries is introduced as a nature-inspired phenomenon. As humans are a part of nature, their social behaviors as their migration should somehow be associated with the physical laws of nature. Human migration is a phenomenon depending on wealth distribution across countries. Wealth as a part of nature tends to have a uniform distribution all over the world. From a physical perspective, it can be distributed by both the conduction phenomenon without human displacement and by convection phenomenon through the movement of humans. In this study, a unique law is expressed for wealth conduction among countries in a mathematical language. The transfer of the wealth between two locations in the world is a function of the equivalent weighted wealth difference between them as well as the conductivity and convective parameter. Wealth as a society characteristic should involve all the important factors as safety, environment, social, political, educational, etc., known as an equivalent weighted wealth of a specific country. Conductivity and convective should cover the factors that exist between the countries, which can facilitate or harden wealth transmission, as the governing government laws, political strategies, and the available distance between the countries. The occurrence of critical conditions naturally or deliberately as war, earthquake, insecurity, etc., which can work as a wealth pump, disturb the wealth distribution by preventing or inversing the wealth transition in the world. This human cancer-like conduction behavior causes wealth to accumulate in wealthy countries. With an increment of wealth between different countries, the human's uncommon convection behavior occurs that leads to the so-called immigration phenomenon. In the migration phenomenon, humans migrate carrying wealth with themselves to wealthy countries. This migration can be illegal until the wealth difference does not exceed the threshold value. In this case, the world will counter will illegal immigrants flowing to wealthy countries through various illegal ways.

**1.4 Methodology of the new approach**
In this paper, we have introduced the equations required for describing the physical behavior of a natural phenomenon. A constitutive relation similar to that of relations applied in physical problems as conduction heat transfer has been presented to show the wealth flow rate among countries. The equivalent wealth of a country as a social characteristic is defined by the factors concerning its value. The conductivity parameter as a wealth movement facilitating or hardening factor is presented with the variables involved. The equation governing the movement of people between locations is introduced to show the flow rate of humans displacing among countries. The connectivity parameter is presented which plays an important role in displacing humans among various locations. Some critique of humans' abnormal behaviors has discussed the outbreak in the face of critical conditions. These atypical behaviors can disturb the distribution of wealth between the countries and lead to illegal immigration eventually. Next, by considering a country as a lumped control volume, the conservation equation of the country's wealth has been presented. By calculating the migration rate of people in the country and considering the birth and death rate of humans, the balanced equation of population dealing with a specific country at any time is also expressed.

**2. Theoretical treatment**
In this section, the physical behavior of a phenomenon occurring in nature is described. Then equations that illustrate the behavior of a natural phenomenon are presented in a mathematical language. Then the wealth distribution in the world as the main factor of human movement is modeled based on the physical laws of nature. The flow rate of human displacement among countries is introduced as a function of the wealth difference and convective parameter. Although humans are a part of nature and should obey the natural physical laws, there are critiques to the nature of human social behavior that is contrary to the equations governing the physical laws of nature.

**2.1. The physical behavior of a phenomenon in nature**
To describe the physical behavior of a phenomenon in nature mathematically, three types of equations are required to be used which are called the balance law, constitutive equation, and kinematic relations. The balance laws and the constitutive equations are always essential in describing the behavior of a phenomenon, but the use of kinematic relations is not always necessary and depending on the type of the phenomena, the kinematic relations can be applied or not.

**2.2 Governing equations in a physical problem**

**2.2.1 Constitutive equations**
In physics, a constitutive equation represents a relationship between two physical quantities that are specific to a substance or material. It indicates the response of the material to external stimuli as applied fields. A constitutive equation is usually expressed in the following form (Incropera, and Dewitt 1985; Halliday 1960)

$$q = -k\nabla\varphi \tag{1}$$

In equation (1), $q$ indicates the flow of a quality that occurs when there is a gradient of the potential $\varphi$. $k$ is called the conductivity parameter and plays an essential role in the flow of the quality. If for example $\varphi = V$ where $V$ is the voltage, $\nabla V$ will represent the difference of voltage across two points. By considering $k = \frac{1}{R}$ where $R$ is the resistance of the conductor, then the $q$ will show the electrical current through the conductor between two points leading to Ohm's law. As another example, if in equation (1) $\varphi = T$ where T is the temperature, $\nabla T$ will show the temperature difference across two points. By considering $k$ the thermal conductivity that depends on the material placed between two points, then, $q$ will show the heat flux through the thermal conductor between two points resulting in Fourier's law. Further, if we assume $\varphi = C$ where $C$ is the mass concentration, $\nabla C$ will be the driving force concentration difference. By considering $k$ the mass transfer coefficient, then $q$ will show the mass transfer between two points with different mass concentrations leading to Fick's law.

### 2.2.2. Conservation laws

In physics, some fundamental laws of continuum mechanics are expressions of the conservation of some physical quantity. These expressions, which are considered the balance or conservation laws, apply to all material continua and result in equations that must always be satisfied. Conservation laws are essential to our understanding of the physical world. They explain which processes can or cannot occur in nature. For example, the conservation law of energy states that the total quantity of energy in an isolated system does not change, though it may change form. In general, the total quantity of property governed by that law remains unchanged during physical processes. Conservation laws are considered to be fundamental laws of nature, with various applications in physics, as well as in other fields such as chemistry, biology, geology, and engineering along with social sciences. Conservation laws dealing with mass, linear, and angular momentum and energy are examples of conservation or balance laws. Particularly, a conservation equation in mathematical language can be expressed in the following form (Pritchard 2011, Zhao 2020)

$$\frac{\partial \vec{V}}{\partial t} + \vec{\nabla}.\vec{J}(\vec{V}) = 0 \tag{2}$$

where, $\vec{V}$ is the conserved vector quantity, $\vec{\nabla}$ denotes the gradient operator, and $\vec{J}$ indicates to the Jacobian of the conserved vector quantity. The equation (2) can be expressed for mass conservation equation as:

$$\frac{\partial \rho}{\partial t} + \vec{\nabla}.(\rho \vec{V}) = 0 \tag{3}$$

where $\rho$ denotes the density of the material

## 2.3. Equations governing wealth distribution among countries

As mentioned in the introduction section, since humans are a part of nature, their behaviors, as migration should somehow obey the physical laws governing natural phenomena. Since this phenomenon is affected by wealth distribution in the world, the definition of a constitutive equation for the wealth conduction similar to that of physics is required (Volpert et al. 2017) Naturally, wealth transition can occur through the conduction phenomenon without human movement or through convection phenomenon via human displacements. In the defined constitutive relation, the potential should cover all factors dealing with the wealth of a country as environment, education, safety, social, which is defined as *equivalent weighted wealth value* of the country. Conductivity and convective should consist of all factors dealing with the wealth convection and conduction between countries. These factors can represent the distance between the countries, government laws, political strategies, etc. In this paper, the following constitutive relation is defined for the wealth conduction from country $j$ to country $i$ as:

$$q_{ij} = (q_{con})_{ij} + (q_{cov})_{ij} = k_{ij}\left((eww)_j - (eww)_i\right) + h_{ij}\left((eww)_j - (eww)_i\right) \tag{4}$$

In equation (4), $q_{ij}$ refers to the total wealth flow rate to the country $i$ from country $j$ $(Q_{ij})$ per unit population of the country $i$ $(P_i)$. $(ewp)_i$ is the potential of the country $i$, which is defined as the equivalent weighted poverty value of country $i$ $(EWW)_i$ per unit of its population $(P_i)$, and $(ewp)_j$ denotes the potential of the country $j$, and is defined as the equivalent weighted poverty value of country $j$ $(EWW)_j$ per unit of its population $(P_j)$. $(q_{con})_{ij}$ is the wealth flow rate to the country $i$ from the country $j$ via wealth conduction. $(q_{cov})$ represents the wealth flow rate to the country $i$ from the country $j$ via wealth convection through human displacement.

The global multidimensional poverty index (MPI) is an international measure of acute poverty in a country. In calculating the MPI for a country, dimensions as health, education, and living standards are used with their specific weights (MPI 2019). In the immigration phenomenon, in addition to the indicators mentioned in the MPI calculation, other factors are also involved. For example, cultural, social, and political poverties not mentioned in the MPI indicators can also be the important factors that may involve in people's immigration. Thus, the MPI of the countries cannot indicate a perfect poverty potential of a country in the migration phenomenon.

The equivalent weighted wealth value of a country consists of all important factors that are involved in the definition of the wealth of that country somehow deal with human life and its welfare. These factors for any country $i$ are defined in the form of the combination of the essential factors as:

$$(EWW)_i = \alpha_i g_i + \beta_i s_i + \kappa_i s'_i + \gamma_i e_i + \lambda_i e'_i + \chi_i p_i + \xi_i h_i + \varsigma_i E_i + \zeta_i c_i \tag{5}$$

In equations (5), $g$ refers to the country's gross productivity factor, $s$ and $s'$ denote the safety and social factors, $e$ and $e'$ represent environmental and economic factors, $p$ shows the political factor, $h_i$ refers to the healthy factor, $E$ and $c$ represent educational and cultural factors. $\alpha, \beta, \gamma, \lambda, \kappa, \varsigma, \xi, \zeta$ are weight coefficients that can vary from one country to another and can be found using machine learning techniques. Evaluation of the mentioned parameters needs professional studies.

In equation (4), $k_{ij}$ is a wealth conductivity coefficient from country $j$ to country $i$ and involves the factors that exist between the two countries. The migration conductivity coefficient is defined as:

$$k_{ij} = f(R_{ij}, G_{ij}) \tag{6}$$

In equation (6), $R$ is a function of the available distance between two locations, and $G$ is a function of government laws and strategies governing relations between two countries. In equation (3), $h_{ij}$ is a wealth connectivity coefficient from country $j$ to country $i$ and involves the factors similar to conductivity parameter as:

$$h_{ij} = \begin{cases} f(R_{ij}, G_{ij}) & ((EWW)_i - (EWW)_j) < (EWW)^* \\ f'(R_{ij}) & ((EWW)_i - (EWW)_j) \geq (EWW)^* \end{cases} \tag{7}$$

In equation (7), $(EWW)^*$ is defined as the threshold weighted wealth value. When the wealth difference between countries exceeds this value, illegal immigration occurs in the world.

In the natural heat, mass, electricity conduction, when there is potential because of characteristic difference, the system tends to make a uniform distribution of the characters in a domain by movement from higher values to the lower ones. For the difference between the nature of wealth distribution and the physical phenomena to be more understandable, the conduction and convection heat transfer as physical phenomena are compared with the wealth distribution in the world. In conduction heat transfer, the temperature is considered as a potential where any difference between two points of material will cause the heat to transfer from the point with a higher temperature to the point with a lower temperature without displacement of molecules. In convection heat transfer, the substance molecules with a high temperature move to the cold point and, the substance molecules with a low temperature move to the point with high temperature. This action will lead to temperature balance across the entire material.

The same phenomenon can occur with wealth or poverty as a social characteristic. It means that in nature, wealth tends to be distributed uniformly in societies. For example, when wealth is accumulated in a special society due to an increase in the production rate, the level of life rises, which leads the production costs to increase. Naturally, the rate of production is starting to decrease and new investment in a poor society is started. This is the natural wealth transfer from wealthy societies to poor ones. This transfer can be done via both the conduction method (without human displacement) and the convection method (with human displacement)

## 2.4. Equations governing human migration between countries

The constitutive equation for the migration flow rate of humans as a convection phenomenon can be expressed as:

$$q_{ij} = h_{ij} \left( (eww)_j - (eww)_i \right) \tag{8}$$

In equation (8), $q_{ij} = \dfrac{Q_{ij}}{P_i}$ refers to the human migration rate to the country $i$ from country $j$ $(Q_{ij})$. $h_{ij}$ is a wealth convective coefficient from the country $j$ to the country $i$ and involves the factors that exist between the two countries and is similar to the conductivity parameter.

## 3. Results and discussions

This paper introduces only a new perspective on migration and is only a new idea. Determining the factors mentioned in theoretical treatment is required to be studied more professionally, so the results conveying statistics are not provided in this study. However, some important results are extracted from the perspective that is mentioned in the following:

### 3.1 Critique of nature human social behavior

Humans as a part of nature tend to show a natural behavior in a way that the world reaches a uniform wealth distribution. Some naturally or deliberately arisen conditions in some countries as earthquakes, war, insecurity, economic instability prepare the condition for the outbreak of human abnormal behaviors. The mentioned conditions work as a heat pump that applies to the societies that cause the natural wealth transmission to prevent or make the wealth transmission direction inverse and it makes some societies wealthier and some poorer. This is called the abnormal human conduction phenomenon. This leads to the intensification of non-uniformity in wealth distribution by accumulating wealth in certain wealthy countries.

By increment of wealth difference between countries the anomalous human convection phenomena called human migration occurs in which human migrates to wealthy countries carrying wealth with themselves. This leads to the legal immigration of humans until the equivalent wealth difference between countries does not exceed the threshold one. If the difference between the equivalent weighted wealth of two countries exceeds the threshold value, the conductivity will only just be a function of the distances between the countries, therefore the wealthier country will counter the mass illegal immigration and the world will counter with a flood of immigrants overflowing to wealthy counties through various illegal ways. This behavior disturbs the natural equilibrium and causes several difficulties that nowadays many countries cope with them.

3.2 Balance equations of population and wealth

For presenting the conservation equation of the population in a specific country, the country $i$ $(C_i)$ is considered as a lamped control volume where the displacement of people within the country is not taken into account. It is shown in Figure 1. It has supposed $N$ countries $(C_1, C_2, C_3, \ldots C_N)$ with a population of $(P_1, P_2, P_3, \ldots P_N)$ at the time $t_0$. $b_i$, the birth rate per population of the country $i$ per unit time has been considered as a population source and $d_i$, the death rate per population of the country $i$ per unit time as a population sink. By considering the population source and sink and expressing the immigration rate of people from $i-1$ countries to the country $i$ at a time interval $\Delta t$, the population of the country $i$ at any time $t$ can be obtained as:

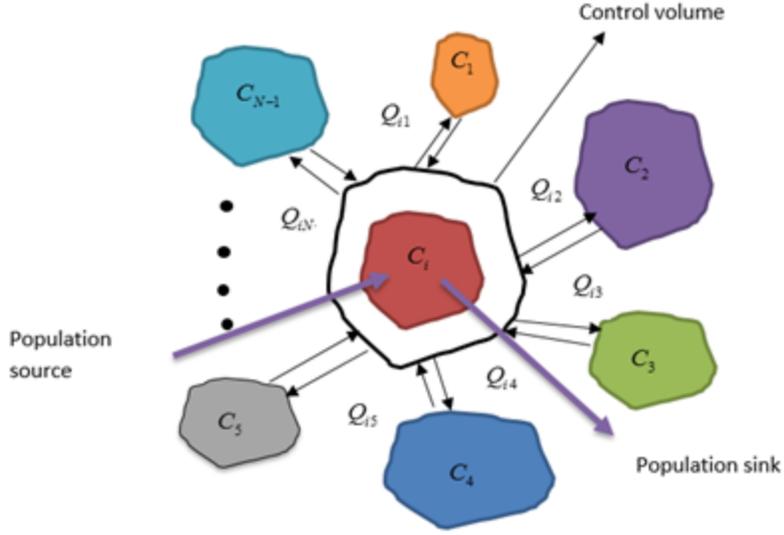

Figure 1. Considering the country $i$ as a control volume

$$P_i(t) \cong P_i(t_0) + P_i b_i \Delta t - P_i d_i \Delta t + \sum_{j=1}^{N} h_{ij} \left( \frac{(EWW)_j}{P_i} - \frac{(EWW)_i}{P_j} \right) P_i \quad i,j=1,....N; \quad k_{ii}=0 \quad (9)$$

For wealth distribution as a combination of conduction and convection phenomenon, the balanced equation of wealth in a specific country can be obtained. By considering $(EWW_p)_i$, the equivalent wealth value production rate per population of the country $i$ per unit time as a wealth source and $(EWW_l)_i$ the equivalent wealth value loss rate per population of the country $i$ per unit time as a wealth sink, the following balance equation can be expressed as:

$$(EWW)_i(t) \cong (EWW)_i(t_0) + P_i(EWW_p)_i \Delta t - P_i(EWW_l)_i \Delta t + \sum_{j=1}^{N} k_{ij} \left( \frac{(EWW)_j}{P_i} - \frac{(EWW)_i}{P_j} \right) P_i$$
$$+ \sum_{j=1}^{N} h_{ij} \left( \frac{(EWW)_j}{P_i} - \frac{(EWW)_i}{P_j} \right) P_i \quad (10)$$

### 4. Conclusion

This paper introduced a perspective for human migration. In this view, inspired by nature, wealth distribution and human migration were modeled based on natural physical laws. The equations and relations governing the behavior of a physical phenomenon were introduced. Some examples of constitutive equations were mentioned describing the behaviors of the physical phenomena in mathematical language as Ohm's law in electricity, Fourier's law in heat transfer, and Fick's law in chemistry. Next, a constitutive equation was proposed for the wealth distribution in the world where the wealth transmission was a function of a potential difference between the countries and a conductivity factor. The potential was defined as equivalent weighted wealth value of country per its population, which was composed of all factors in a country that could affect the human's welfare and life as the country gross productivity factor, environmental, educational, social, and safety factors. In the proposed constitutive relation for wealth conductivity, the conductivity included the factors of available distance and the government laws governing between the countries. It was shown that due to some conditions that arise in countries as war, insecurity, etc. which work as a wealth pump, the wealth distribution is disturbed. This abnormal conduction behavior of humans inverses the wealth transmission in the world and causes the wealth to be accumulated in wealthy countries. This huge difference in the wealth and poverty distribution causes the human to show uncommon convection behavior which leads to the occurrence of migration phenomenon to occur. In such a condition when the difference of equivalent weighted poverty wealth between countries exceeds the threshold value, the conductivity will only be a function of the distance between locations and the displacement of people will not obey the natural laws. Thus, the world will encounter the illegal migration of humans. It is concluded that maintenance of the equivalent weighted poverty between the countries lower than the threshold value will help reduce illegal immigration in the world. The introduced nature-inspired model can be useful for researchers, leaders and rulers.

**The message of the Paper**
*This paper tries to deliver a message to wealthy countries that if the equivalent weighted wealth value between the wealthy countries and the others exceeds the threshold value, the humans' abnormal behaviors will emerge, and therefore they will counter a flood of immigrants overflowing to those counties from various illegal ways.*

*Authors have no conflict of interest to declare*
*The authors received no financial support for the research, authorship, and/or publication of this article.*